\documentclass[12pt]{article}
\usepackage{latexsym}
\usepackage{graphicx}
\usepackage{caption}
\usepackage{psfrag}
\usepackage{amsmath,amssymb,dsfont}
\newcommand{\be}{\begin{equation}}
\newcommand{\ee}{\end{equation}}
\newcommand{\bea}{\begin{eqnarray}}
\newcommand{\eea}{\end{eqnarray}}

\newcommand{\cO}{{\cal O}}
\newcommand{\Frac}[2]{\frac{\displaystyle #1}{\displaystyle #2}}

\input epsf

\begin{document}

\begin{titlepage}

\begin{flushright}
%\today
\end{flushright}
\vspace*{1.5cm}
\begin{center}
{\Large \bf Vector Meson Dominance\\
as a first step in a systematic approximation: \\
\vspace*{0.3cm}
the pion vector form factor
}\\[3.0cm]

{\bf P. Masjuan, S. Peris} and {\bf J.J. Sanz-Cillero}\\[1cm]

Grup de F{\'\i}sica Te{\`o}rica and IFAE\\ Universitat Aut{\`o}noma de Barcelona, 08193 Barcelona, Spain.\\[0.5cm]

\end{center}

\vspace*{1.0cm}

\begin{abstract}

Pad\'{e} Approximants can be used  to go beyond  Vector Meson Dominance in a systematic approximation. We
illustrate this fact with the case of the  pion vector form factor and extract values for the first two
coefficients of its Taylor expansion.  Pad\'e Approximants are shown to be a useful and simple tool for
incorporating high-energy information, allowing an improved determination of these Taylor coefficients.
\end{abstract}

\end{titlepage}

\section{Introduction}

It has been known for a long time that the pion vector form factor (VFF) in the space-like region is very
well described by a monopole ansatz of the type given by Vector Meson Dominance (VMD) in terms of the rho
meson. However, it has remained unclear whether there is a good reason for this from QCD or it is just  a
mere coincidence  and,  consequently, it is not known how to go about improving on this ansatz.

To begin our discussion, let us define  the form factor, $ F(Q^2)$, by the matrix element
\begin{equation}\label{def}
    \langle \pi^{+}(p')| \ \frac{2}{3}\ \overline{u}\gamma^{\mu}u-\frac{1}{3}\ \overline{d}\gamma^{\mu} d-
    \frac{1}{3}\ \overline{s}\gamma^{\mu} s\ | \pi^{+}(p)\rangle= (p+p')^{\mu} \ F(Q^2)\ ,
\end{equation}
where $Q^2=-(p'-p)^2$, such that $Q^2>0$ corresponds to space-like data.
Since  the spectral function for the corresponding dispersive integral for $F(Q^2)$
starts at twice the pion mass, the form factor can be approximated by a Taylor expansion
in powers of the momentum for $|Q^2|< (2 m_\pi)^2$.
%%, i.e.
%%
%%\begin{equation}\label{Taylor}
%%    F(Q^2)\approx 1\ - \ \frac{1}{6}\ \langle r^2\rangle\ Q^2 \ +\ c \ Q^4 \ +\ d \ Q^6 \ ...
%%\end{equation}
%%
At low momentum, Chiral Perturbation Theory is the best tool for organizing the pion interaction in a
systematic expansion in powers of momenta and quark masses~\cite{chpt-Weinberg,chpt-SU2,chpt-SU3}. With
every order in the expansion, there comes a new set of coupling constants, the so-called low-energy
constants (LECs), which encode all the QCD physics from higher energies. This means, in particular, that the
coefficients in the Taylor expansion  can be expressed in terms of these LECs and powers of the quark
masses. Consequently, by learning about the low-energy expansion, one may indirectly  extract important
information about QCD.

In principle, the coefficients in the Taylor expansion
may be obtained by means of a polynomial fit to the experimental data
in the space-like region~\footnote{Time-like data is provided
by $\pi\pi$ production experiments and, consequently,
they necessarily correspond to values of the momentum  above the $\pi\pi$ cut, i.e. $ |Q^2|> 4m_\pi^2$ with $Q^2<0$.}
below $Q^2=4m_{\pi}^2$. However, such a polynomial fit
implies a tradeoff. Although,
 in order to decrease  the (systematic) error of the truncated Taylor expansion,
 it is clearly better to go to a low-momentum region, this also downsizes the set
 of data points  included in the fit which, in turn, increases the (statistical) error.
In order to achieve a smaller statistical error one would have to include experimental data from higher energies, i.e. from $Q^2> 4 m_\pi^2$.  Since this is not possible
in a polynomial fit, the use of alternative mathematical descriptions
may be a better strategy.

One such description, which includes time-like data as well, is based on the use of the Roy equations and
Omn\'es dispersion relations. This is the avenue  followed by \cite{Colangelo,ColangeloB}, which has already
produced interesting results on the scalar channel \cite{Caprini}, and which can also be applied to the
vector channel. Other procedures have relied on conformal transformations for the joint analysis of both
time-like and space-like data~\cite{Yndurain},  or  subtracted Omn\'es relations~\cite{Pich, Portoles}. Further analyses may be found in Ref. \cite{Caprini2}.

On the other hand, as already mentioned above, one may also consider an ansatz of the type
\begin{equation}\label{vmd}
    F(Q^2)_{_{\rm VMD}}=\left(1+\frac{Q^2}{M^2_{_{\rm VMD}} }\right)^{-1}\ .
\end{equation}
Even though the simplicity of the form of Eq.~(\ref{vmd}) is quite astonishing, it reproduces the space-like
data rather well, even for a range of momentum of the order of a few GeV, i.e. $Q^2\gg 4 m_\pi^2$.  If this
fact is not merely a fluke, it could certainly be interesting to consider the form (\ref{vmd}) as the first
step in a systematic approximation,  which would then allow improvement on this VMD ansatz.

In this article, we would like to point out that the previous VMD ansatz for the  form factor (\ref{vmd})
can be viewed as the first element in a sequence of Pad\'{e} Approximants (PAs) which can be constructed in a
systematic way. By considering higher-order  terms in the  sequence, one may be able to describe the
space-like data with an increasing level of accuracy~\footnote{Obviously, unlike the space-like data, one
should not expect to reproduce the time-like data since a Pad\'e Approximant contains only isolated poles
and cannot reproduce a time-like cut.}. Of course, whether this is actually the case  and the sequence is a
convergent one in the strict mathematical sense or, on the contrary, the sequence eventually diverges,
remains to be seen. But the important difference with respect to the traditional VMD approach is that, as a
Pad\'e sequence, the approximation is well-defined and can be systematically improved upon.

Although polynomial fitting is more common, in general, rational approximants (i.e. ratios of two
polynomials) are able to approximate the original function in a much broader range in momentum than a
polynomial~\cite{Baker}. This will be the great advantage of the Pad\'es compared to other methods: they
allow the inclusion of low and high energy information in a rather simple way which, furthermore, can in
principle be systematically improved upon. In certain cases, like when the form factor obeys a dispersion
relation given in terms of a positive definite spectral function (i.e. becomes a Stieltjes function), it is
known that the Pad\'e sequence is convergent everywhere on the complex plane, except on the physical cut
\cite{PerisPades}. Another case of particular interest is in the limit of an infinite number of colors in
which the form factor becomes a  meromorphic function. In this case there is also a theorem which guarantees
convergence of
 the Pad\'{e} sequence everywhere in a compact region of the complex plane, except perhaps at a
 finite number of points (which include the poles in the spectrum contained in that region)
 \cite{PerisMasjuan07}.
In the real world, in which a general form factor has a complicated analytic structure with a cut, and whose
spectral function is not positive definite, we do not know of any mathematical result assuring the
convergence of a Pad\'{e} sequence \cite{JuanjoVirtoMasjuan}. One just has to try the approximation on the data
to learn what happens.

In this work we have found that, to the precision allowed by the experimental data, there are sequences of
PAs which improve on the lowest order VMD result in a rather systematic way. This has allowed us to extract
the values of the lowest-order coefficients of the low-energy  expansion.

We would like to emphasize that, strictly speaking, the Pad\'e Approximants
to a given function are ratios  of two polynomials $P_N(z)$ and $Q_M(z)$
(with degree $N$ and $M$, respectively), constructed such that the Taylor expansion
around the origin exactly coincides with that of $f(z)$ up to the highest possible order, i.e.
$f(z)-R^N_M(z) =\cO(z^{M+N+1})$. However, in our case the Taylor coefficients
are not known. They are, in fact, the information we are seeking for.
Our strategy will consist in determining
these  coefficients by a least-squares fit of a Pad\'e Approximant
to the vector form factor data in the space-like region.

There are several types of PAs that may be considered. In order to achieve a fast numerical convergence, the
choice of which one to use is largely determined by the analytic properties of the function to be
approximated. In this regard, a glance at the time-like data of the pion form factor makes it obvious that
the form factor is clearly dominated by the rho meson contribution. The effect of higher resonance states,
although present, is much more suppressed. In these circumstances the natural choice is a $P^{L}_{1}$ Pad\'{e}
sequence~\cite{Baker}, i.e. the ratio of a polynomial of degree $L$ over a polynomial of degree
one~\footnote{Conventionally, without loss of generality, the polynomial in the denominator is normalized to
unity at the origin.}. Notice that, from this perspective, the VMD ansatz in (\ref{vmd}) is nothing but the
$P^0_1$ Pad\'{e} Approximant.

However, to test the aforementioned single-pole dominance, one should check the degree to which the
contribution from resonances other than the rho may be neglected. Consequently, we have also considered the
sequence $P^{L}_{2}$, and the results confirm those found with the PAs $P^{L}_{1}$. Furthermore, for
completeness, we have also considered the so-called Pad\'e-Type approximants (PTs)~\cite{math,PerisMasjuan08}. These are
rational approximants whose poles are predetermined at some fixed values, which we take to be the physical
masses since they are known. Notice that this is different from the case of the ordinary PAs, whose poles
are left free and determined by the fit. Finally, we have also considered an intermediate case, the
so-called Partial-Pad\'e approximants (PPs)~\cite{math}, in which some of the poles are predetermined (again
at the physical masses) and some are left free. We have fitted all these versions of rational approximants
to all the available pion VFF space-like data~\cite{Amendolia}-\cite{Dally}. The result of
the fit is rather independent of the kind of rational approximant sequence used and all the results show
consistency among themselves.

The structure of this letter is the following.
In section~\ref{sec:model}
we begin by testing the efficiency of the $P^L_1$ Pad\'es with the help of a model.
In section \ref{sec:FF} we apply this very same method  to the
experimental VFF. Firstly, in sec.~\ref{sec:PAL1}, we use the
Pad\'e Approximants $P^L_1$; then, in Sec.~\ref{sec:PAL2}, this result
is cross-checked with a $P^L_2$ PA. Finally,  in sec.~\ref{sec:PTPP}, we study the
Pad\'e-Type and Partial-Pad\'e approximants.
The outcome of these analyses is combined in section~\ref{sec:results}
and some conclusions are extracted.

\section{A warm-up model}\label{sec:model}

In order to illustrate the usefulness of the PAs as fitting functions in the way we propose here,
we will first use a phenomenological model as a theoretical laboratory to check our method. Furthermore, the model will also give us an idea about the size of possible systematic uncertainties.

We will consider a VFF phase-shift with the right threshold behavior
and with approximately the physical values of the rho mass and width. The form-factor
is recovered through a once-subtracted Omn\'es relation,
\begin{equation}\label{model}
    F(Q^2)=\exp\left \{-\frac{Q^2}{\pi} \int_{4 \hat{m}_{\pi}^{2}}^{\infty}\ dt\ \frac{\delta(t)}{t (t+Q^2)}\right\}\ ,
\end{equation}
where $\delta(t)$, which plays the role of the vector form factor phase-shift~\cite{Pich, Portoles, Cillero}, is given by
\begin{equation}\label{model2}
    \delta(t)=\tan^{-1}\left[\frac{\hat{M}_{\rho}
    \hat{\Gamma}_{\rho}(t)}{\hat{M}_{\rho}^2-t} \right]\ ,
\end{equation}
with the $t$-dependent width given by
\begin{equation}\label{width}
    \hat{\Gamma}_{\rho}(t)= \Gamma_{0}\ \left( \frac{t}{\hat{M}_{\rho}^2} \right)\ \frac{\sigma^3(t)}{\sigma^3(\hat{M}_{\rho}^2)}\ \theta\left( t- 4 \hat{m}_{\pi}^{2} \right)\ ,
\end{equation}
and $\sigma(t)=\sqrt{1-4 \hat{m}_{\pi}^{2}/t}$.
The input parameters  are chosen to be close to their physical values:
\begin{equation}\label{param}
 \Gamma_{0} = 0.15\ \mathrm{GeV}\quad ,\quad
 \hat{M_{\rho}}^2= 0.6\ \mathrm{GeV}^2\quad ,\quad
 4 \hat{m}_{\pi}^{2}= 0.1 \ \mathrm{GeV}^2\, .
\end{equation}
 We emphasize that the model defined by the expressions (\ref{model}-\ref{width}) should be considered
as quite realistic. In fact, it has been used in Ref.~\cite{Pich, Portoles, Cillero} for extracting the values for the
physical mass and width of the rho meson through a direct fit to the   (timelike) experimental data.

Expanding $F(Q^2)$ in Eq. (\ref{model}) in powers of $Q^2$ we readily obtain
\begin{equation}\label{expmodel}
    F(Q^2)%%\approx
    \, =\, 1 \, - \,  a_1\ Q^2 \, + \,  a_2\ Q^4 \,  - \ a_3\ Q^6 + ... \,\, ,
\end{equation}
with known values for the coefficients $a_i$. In what follows, we will use Eq. (\ref{expmodel}) as the
definition of the coefficients $a_i$. To try to recreate the situation of the experimental
data~\cite{Amendolia}-\cite{Dally} with the model, we have generated fifty ``data'' points
in the region $0.01\leq Q^2\leq 0.25$, thirty data points in the interval $0.25\leq Q^2 \leq 3$, and seven
points for $3\leq Q^2\leq 10$ (all these momenta in units of GeV$^2$). These points are taken with vanishing
error bars since  our purpose here is  to estimate the systematic error derived purely from our approximate
description of the form factor.

We have fitted a sequence of Pad\'e Approximants $P^{L}_{1}(Q^2)$ to these data points and, upon expansion
of the Pad\'{e}s around $Q^2=0$, we have used them to predict the values of the coefficients $a_i$. The
comparison may be found in Table~\ref{table1}. The last PA we have fitted to these data is $P^6_1$. Notice
that the pole position of the Pad\'{e}s differs from the true mass of the model, given in Eq.~(\ref{param}).

\begin{table}[t]
\centering
\begin{tabular}{|c|c|c|c|c|c|c|c|c|}
  \hline
   & $P^{0}_{1}$ &  $P^{1}_{1}$ &  $P^{2}_{1}$ &  $P^{3}_{1}$&$ P^{4}_{1}$  & $P^{5}_{1}$  &$P^{6}_{1}$ & $F(Q^2)$(exact)\\ \hline
  $a_1$(GeV$^{-2}$) & 1.549 & 1.615 & 1.639 & 1.651 & 1.660&1.665 & 1.670 & 1.685 \\
  $a_2$ (GeV$^{-4}$)& 2.399 & 2.679& 2.809 & 2.892 & 2.967&3.020 & 3.074& 3.331\\
  $a_3$(GeV$^{-6}$)& 3.717 & 4.444 & 4.823 & 5.097 & 5.368&5.579 & 5.817& 7.898\\
  \hline
  \hline
  $s_p$(GeV$^{2}$) &$0.646$&$0.603$&$0.582$&$0.567$&$0.552$&$0.540$&$0.526$&$0.6$\\
  \hline
\end{tabular}
\caption{{\small Results of the various fits to the form factor $F(Q^2)$ in the model, Eq. (\ref{model}).
The exact values for the coefficients $a_i$ in Eq. (\ref{expmodel}) are given on the last column. The last
row shows the predictions for the corresponding pole for each Pad\'{e} ($s_p$), to be compared to the true mass
$\hat{M}_{\rho}^{2}=0.6\ $GeV$^2$ in the model.}} \label{table1}
\end{table}

A quick look at Table \ref{table1} shows that the sequence seems to converge to the exact result, although
in a hierarchical way, i.e. much faster for $a_1$ than for $a_2$, and this one much faster than $a_3$,
etc... The relative error achieved in determining the coefficients $a_i$ by the last Pad\'e, $P^6_1$, is
respectively $0.9\%$, $8\%$ and $26\%$ for $a_1, a_2$ and $a_3$. Naively, one would expect these results to
improve as the resonance width decreases since the $P^{L}_{1}$ contains only a simple pole, and  this is
indeed what happens. Repeating this exercise with the model, but with a $\Gamma_0=0.015$~GeV ($10$ times
smaller than the previous one), the relative error achieved by $P^6_1$ for the same coefficients as before
is $0.12\%$, $1.1\%$ and $4.7\%$. On the other hand, a model with $\Gamma_0$ five times bigger than the
first one produces, respectively, differences of $2.1\%$, $14.4\%$ and $37.8\%$.

As we have mentioned in the introduction, it is possible to build a variation of the PAs, the Pad\'e-Type
Approximants, where one fixes the pole in the denominator at the physical mass and only the numerator is
fitted. We have also studied the convergence of this kind of rational approximant with the model. Thus, in
this case, we have placed the $P^L_1$ pole at $s_p=\hat{M}^2_\rho$ and found a similar pattern as in
Table~\ref{table1}. For $P^6_1$,  the Pad\'e-Type coefficient $a_1$ differs a $2.5\%$ from its exact value,
$a_2$ by $16\%$ and $a_3$ by $40\%$.

Based on the previous results, we will take the values in  Table \ref{table1} as a rough estimate of the systematic uncertainties when fitting to the experimental data in the following sections. Since, as we will see, the best fit to the experimental data comes from the  Pad\'{e} $P^4_1$, we will take the error in Table \ref{table1} from this Pad\'{e} as a reasonable estimate and add to the final error  an extra systematic uncertainty of $1.5\%$ and $10\%$  for $a_1$ and $a_2$ (respectively).

\section{The pion vector  form factor}\label{sec:FF}

We will use all the available experimental data in the space-like region, which may be found in
Refs.~\cite{Amendolia}-\cite{Dally}.
These data range in momentum from $Q^2=0.015$ up to 10~GeV$^2$.

As discussed in the introduction, the prominent role of the rho meson contribution motivates
that we start with the $P^{L}_{1}$ Pad\'e sequence.

\subsection{Pad\'e Approximants $P^{L}_{1}$}\label{sec:PAL1}

\begin{figure}
  \center
  \includegraphics[width=13cm]{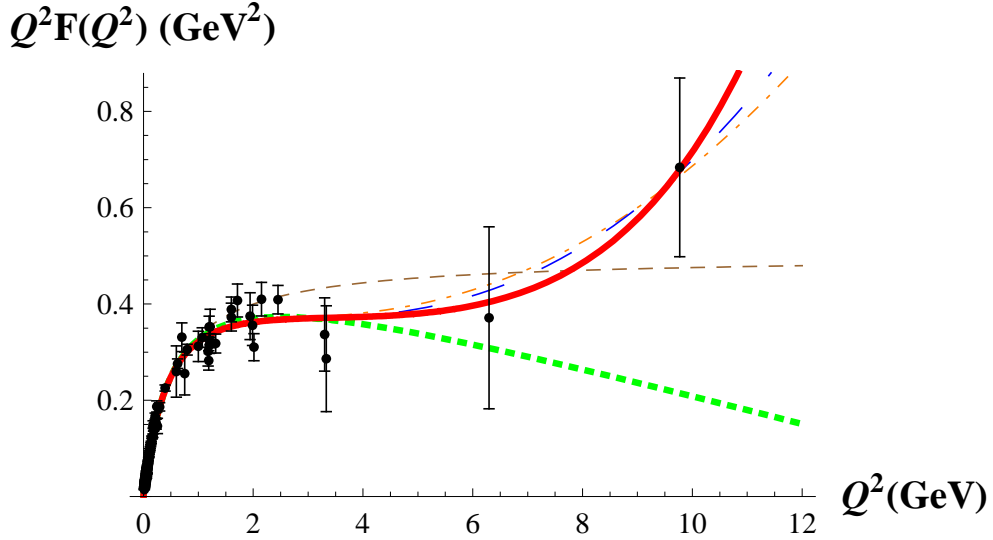}
  \caption{{\small The sequence of $P^L_1$ PAs is compared to the available space-like
  data~\cite{Amendolia}-\cite{Dally}:
  $P^0_1$ (brown dashed), $P^1_1$ (green thick-dashed),
  $P^2_1$ (orange dot-dashed), $P^3_1$ (blue long-dashed),
  $P^4_1$ (red solid).}}\label{fig:VFF}
\end{figure}

Without any loss of generality, a $P^{L}_{1}$ Pad\'e is given by
\begin{equation}
P_1^L(Q^2) \,\, \,= \,\,\,
1\, +\,  \sum_{k=0}^{L-1}a_k (-Q^2)^{k} \,\,
+ \, (-Q^2)^{L} \,  \frac{\ a_L}{1+\Frac{a_{L+1} }{a_L}  \, Q^2}\  ,
\label{PL1}
\end{equation}
where the vector current conservation condition $P^L_1(0)=1$ has been imposed
and the coefficients $a_{k}$ are the low-energy coefficients of the corresponding Taylor expansion of the VFF (compare with (\ref{expmodel}) for the case of the model in the previous section).

The fit of $P^L_1$ to the space-like data points in Refs. ~\cite{Amendolia}-\cite{Dally}
determines  the coefficients  $a_{k}$ that best interpolate them.  According to Ref.~\cite{brodsky-lepage},
the form factor is supposed to fall off as $1/Q^2$ (up to logarithms) at large values of $Q^2$. This means
that, for any value of $L$, one may expect to obtain a good fit only up to a finite value of $Q^2$, but not
for asymptotically large momentum. This is clearly seen in Fig. ~\ref{fig:VFF}, where the Pad\'{e} sequence
$P^L_1$ is compared to the data up to $L=4$.

\begin{figure}[!t]
  \center
  \includegraphics[width=7cm]{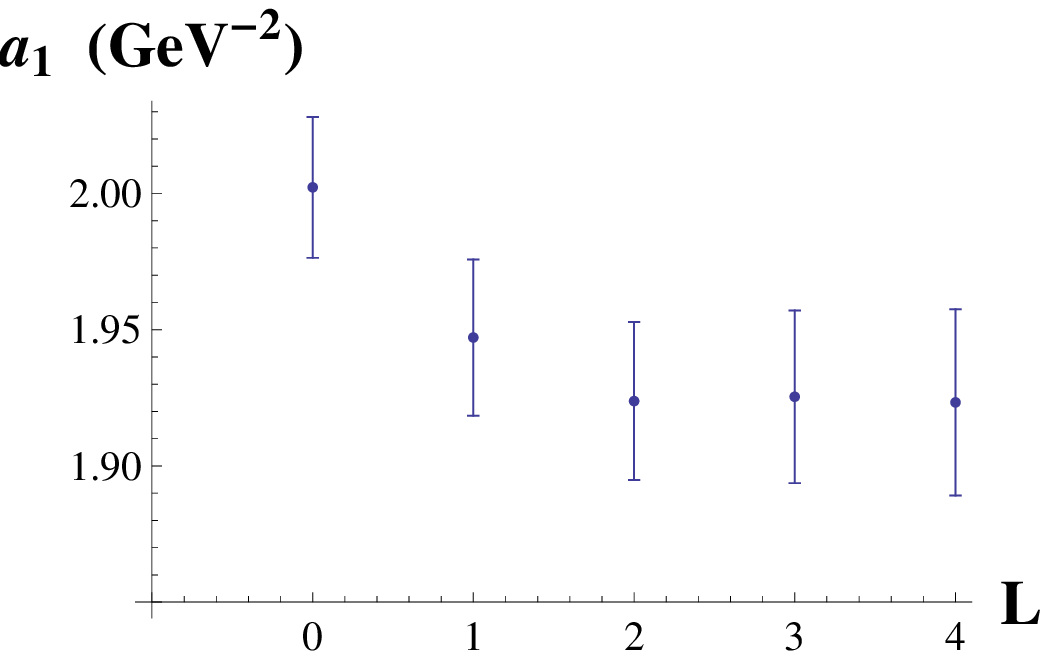}
  \hspace*{1.cm}
  \includegraphics[width=7cm]{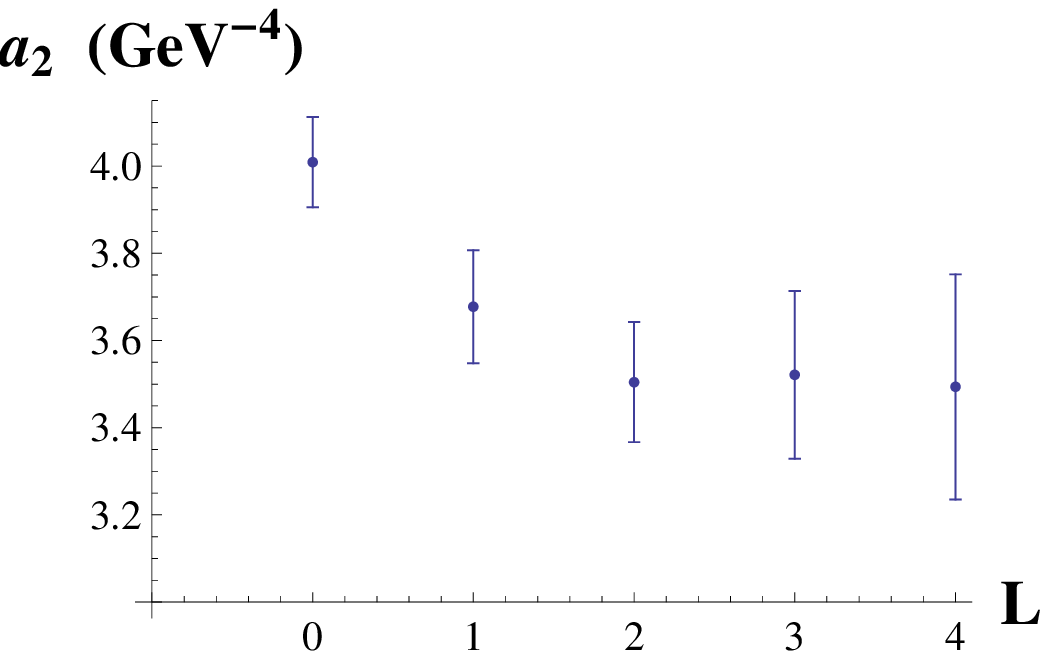}
  \caption{{\small $a_1$ and $a_2$ Taylor coefficients for the
  $P^L_1$ PA sequence.  }}\label{fig:a1PL1}
\end{figure}

Fig.~\ref{fig:a1PL1} shows the evolution of the fit results for the Taylor
coefficients $a_1$ and $a_2$ for the $P^L_1$ PA  from $L=0$ up to $L=4$.
As one can see, after a few Pad\'es
these coefficients become stable. Since the experimental data have non zero error
it is only possible to fit a $P^L_1$ PA up to a certain value for $L$. From this order on, the large error bars in the highest coefficient in the
 numerator polynomial make it compatible with zero and, therefore, it no
 longer makes sense to talk about a new element in the sequence. For the
 data in Refs. ~\cite{Amendolia}-\cite{Dally}, this happened
 at $L=4$ and this is why our plots stop at this value. Therefore, from the PA  $P^4_1$
 we obtain our best fit and, upon expansion around $Q^2=0$,  this yields
\begin{equation}
a_1\, =\, 1.92 \pm 0.03\,\,\mbox{GeV}^{-2} \, ,  \qquad\qquad a_2\, =\, 3.49 \pm 0.26\,\,\mbox{GeV}^{-4} \,
;
\end{equation}
with a $\chi^2/\mathrm{dof}=117/90$.

\begin{figure}
  \center
  \includegraphics[width=6.5cm]{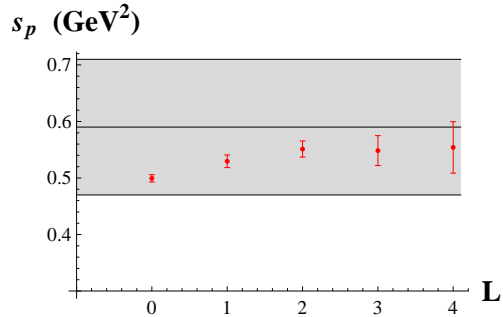}
  \caption{{\small Position $s_p$ of the pole for the different $P^L_1$.
  The range with the physical values
  $M_\rho^2\pm M_\rho\Gamma_\rho$ is shown (gray band)
  for comparison. }}\label{fig:spPL1}
\end{figure}

Eq.~(\ref{PL1}) shows that the pole of each $P^L_1$ PA is determined by the ratio $s_p=a_L/a_{L+1}$.  This
ratio is shown in  Fig.~\ref{fig:spPL1}, together with a gray band whose width is given by $\pm
M_\rho\Gamma_\rho$ for comparison. From this figure
 one can see that the position of the pole of the PA  is close to the
physical value $M_\rho^2$~\cite{PDG}, although it does not necessarily agree with it, as we already saw in the model of the previous section.

\subsection{Comment on $P^L_2$ Pad\'es}  \label{sec:PAL2}

Although the time-like data of the pion form factor
is clearly dominated by the $\rho(770)$ contribution,
 consideration of two-pole
$P^L_2$ PAs will give us a way to assess any possible systematic bias in our previous analysis, which was limited to only single-pole PAs.

We have found that the results of the fits of $P^L_2$ PAs to the data tend to reproduce
the VMD pattern found for the $P^L_1$ PAs in the previous section.  The $P^L_2$ PAs place the first of the two poles around the rho mass, while the second wanders around the complex momentum plane together with a close-by zero in the numerator. This association of a pole and a close-by zero is what is called a ``defect" in the mathematical literature\cite{Baker2}. A defect is only a local perturbation and, at any finite distance from it, its effect is essentially negligible. This has the net effect that the $P^L_2$ Pad\'e in the euclidean region looks just like a $P^L_1$ approximant and, therefore, yields essentially the same results. For example, for the $P^2_2$, one gets
\begin{equation}
a_1\, =\, 1.924 \pm 0.029 \,\,\mbox{GeV}^{-2} \, ,  \qquad\qquad
a_2\, =\, 3.50  \pm 0.14 \,\,\mbox{GeV}^{-4} \,  ,
\end{equation}
with  a $\chi^2/\mathrm{dof}= 120/92$.

\subsection{Pad\'{e} Type and Partial Pad\'e Approximants}\label{sec:PTPP}

Besides the ordinary Pad\'e Approximants one may consider other kinds of rational approximants. These are
the Pad\'{e} Type and Partial Pad\'e Approximants ~\cite{PerisMasjuan07, math, PerisMasjuan08}. In the Pad\'e
Type Approximants (PTAs) the poles of the Pad\'e are fixed to certain particular values, which in this
context are naturally the physical masses.  On the other hand, in the Partial Pad\'e Approximants (PPAs) one
has an intermediate situation beytween the PAs and the PTAs in which some poles are fixed while others are
left as free parameters to fit.

\begin{figure}[!t]
 \center
 \includegraphics[width=8cm]{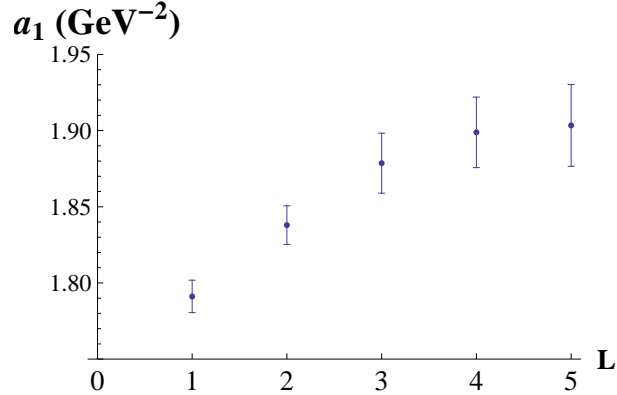}\\
 \caption{Low-energy coefficient $a_1$ from the
 $T^L_1$ Pad\'e-Type sequence}\label{fig:a1PTL1}
\end{figure}

Since the value of the physical rho mass is known ($M_{\rho}=775.5$~MeV), it is natural to attempt a fit of PTAs to the data with a pole fixed at that mass. The corresponding sequence will be called $T^L_1$. This has the obvious advantage that the number of parameters in the fit decreases by one and allows one to go a little further in the sequence.  Our best value is then given
by the Pad\'e Type  Approximant $T^5_1$, whose expansion around $Q^2=0$ yields the following values for the Taylor coefficients:
\begin{equation}
a_1\, =\, 1.90 \pm 0.03\,\,\mbox{GeV}^{-2} \, ,  \qquad\qquad a_2\, =\, 3.28  \pm  0.09 \,\,\mbox{GeV}^{-4}
\, ,
\end{equation}
with a $\chi^2/\mathrm{dof}=118/90$.

The previous analysis of PTAs may be extended by making further use of our knowledge of the vector
spectroscopy~\cite{PDG}. For instance, by taking $M_{\rho}=775.5$~MeV, $M_{\rho'}=1459$~MeV and
$M_{\rho''}=1720$~MeV,\footnote{As will be seen, results do not depend on the precise value chosen for these masses.}  we may construct  further Pad\'e-Type sequences of the form $T^L_2$ and $T^L_3$.

\begin{figure}[!t]
 \center
 \includegraphics[width=7cm]{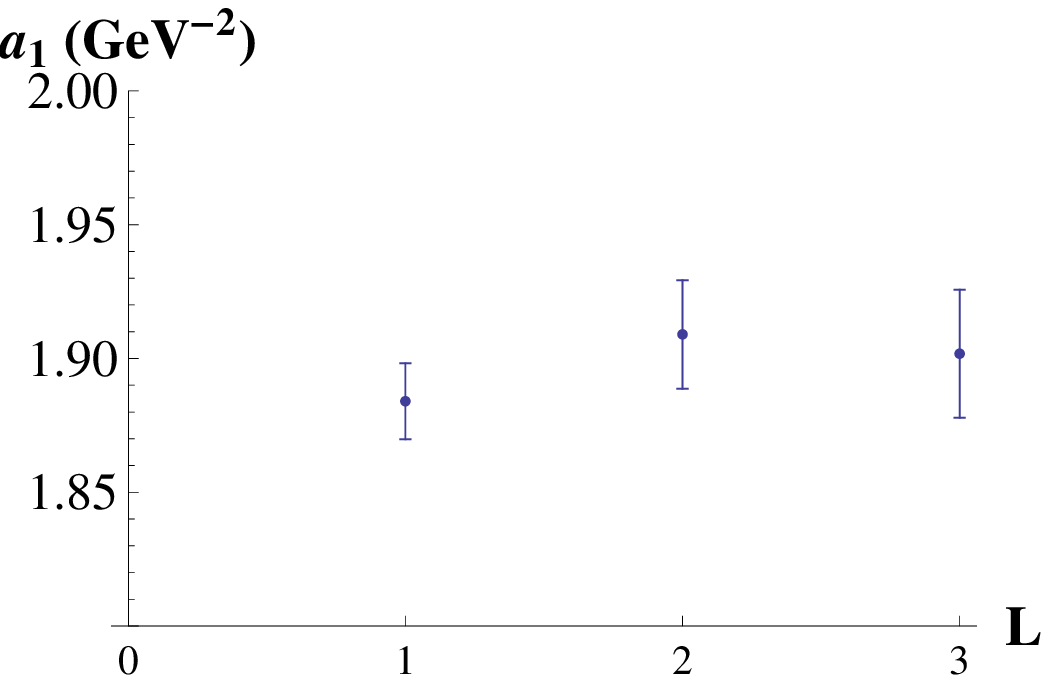}
 \hspace*{.5cm}
 \includegraphics[width=7cm]{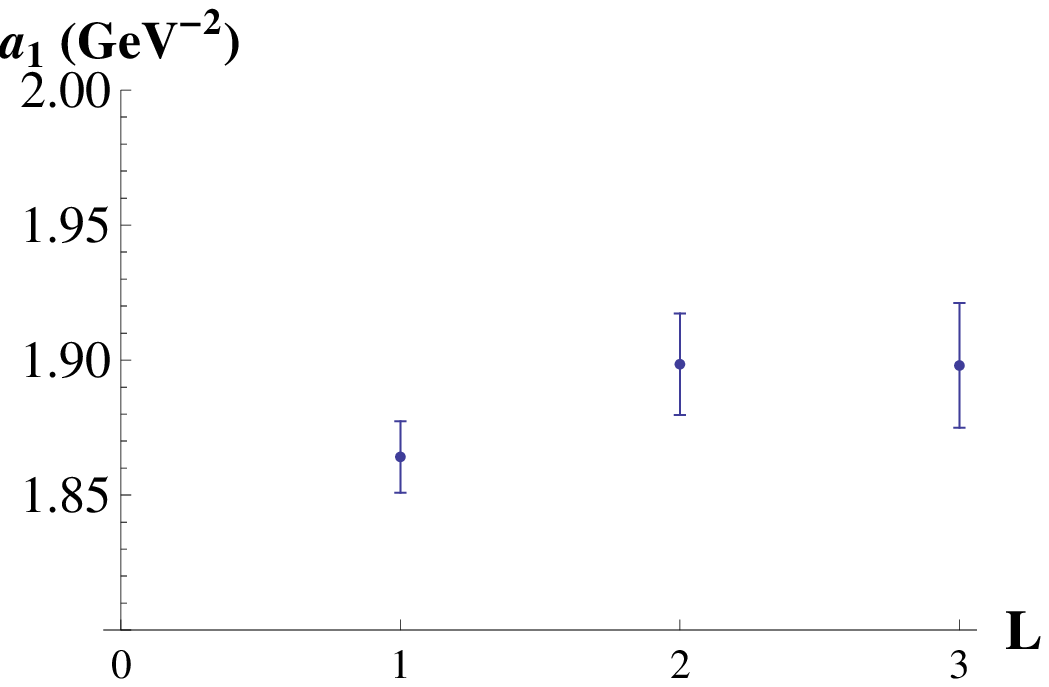}
 \caption{{\small
 Low energy coefficient $a_1$  for the $T^L_2$ Pad\'e-Type
 sequence with $M_{\rho}$ and $M_{\rho'}$ (left),
 and with $M_{\rho}$ and $M_{\rho''}$ (right).
 }}
 \label{fig:a1PT2}
\end{figure}

In the PTA sequence $T^L_2$ one needs to provide the value of two poles. For the first pole, the natural
choice is $M_{\rho}^2$. For the second pole,  we found that choosing either $M_{\rho'}^2$ or $M_{\rho''}^2$
(the second vector excitation) does not make any difference. Both outcomes are compared in
Fig.~(\ref{fig:a1PT2}). Using $M_{\rho'}^2$, we found that the $T^3_2$ PTA yields the best values as
\begin{equation}
a_1\, =\, 1.902 \pm 0.024\,\,\mbox{GeV}^{-2} \, ,  \qquad\qquad a_2\, =\, 3.29  \pm  0.07
\,\,\mbox{GeV}^{-4} \, ,
\end{equation}
with a $\chi^2/\mathrm{dof}=118/92$.

 Using $M_{\rho''}^{2}$ as the second pole  one also gets the best value from the  $T^3_2$ PTA, with
the following results:
\begin{equation}
a_1\, =\, 1.899 \pm 0.023\,\,\mbox{GeV}^{-2} \, ,  \qquad\qquad a_2\, =\,  3.27 \pm 0.06 \,\,\mbox{GeV}^{-4}
\, ,
\end{equation}
with a $\chi^2/\mathrm{dof}=119/92$. We find the stability of the results for the coefficients $a_{1,2}$ quite
reassuring.

We have also performed an analysis of  the PTA sequence $T^L_3$, with similar conclusions. From the $T^3_3$
we obtain  the following values for the coefficients:
\begin{equation}
a_1\, =\, 1.904 \pm 0.023\,\,\mbox{GeV}^{-2} \, ,  \qquad\qquad a_2\, =\, 3.29  \pm 0.09 \,\,\mbox{GeV}^{-4}
\, ,
\end{equation}
with a $\chi^2/\mathrm{dof}=119/92$.

Finally, to complete our analysis, we will also consider Partial Pad\'e Approximants, in which only part of the denominator is given in advance. In particular, we study the PPA sequence $P^L_{1,1}$ \footnote{See Ref. \cite{PerisMasjuan07} for notation.} in which
the first pole is given by $M_{\rho}^2$ and the other is left  free. The best determination of the Taylor coefficients is  given by $P^2_{1,1}$, and they yield
\begin{equation}
a_1\, =\, 1.902  \pm 0.029 \,\,\mbox{GeV}^{-2} \, ,  \qquad\qquad
a_2\, =\, 3.28   \pm 0.09  \,\,\mbox{GeV}^{-4} \, ,
\end{equation}
with  the free pole  of the PPA given by $M_{free}^2=(1.6   \pm 0.4 $~GeV$)^2$ and a $\chi^2/\mathrm{dof}=119/92$.

\section{Combined Results and conclusions}\label{sec:results}
 \label{sec:conclusion}

Combining all the previous rational approximants results
in an average given by
\begin{equation}
a_1\, =\, 1.907\pm 0.010_{\mathrm{stat}} \pm 0.03_{\mathrm{syst}}\,\,\mbox{GeV}^{-2} \  ,  \
a_2\, =\,  3.30 \pm  0.03_{\mathrm{stat}} \pm 0.33_{\mathrm{syst}}\,\,\mbox{GeV}^{-4} \, .
\end{equation}
The first error comes from combining the results of the different fits by means of a weighted average. On
top of that, we have added what we believe to be a conservative estimate of the theoretical (i.e.
systematic) error based on the analysis of the VFF model in Sec.~\ref{sec:model}. We expect the latter to
give an estimate for the systematic uncertainty due to the approximation of the physical form factor with
rational functions. For comparison with previous analyses, we also provide in Table \ref{table2} the value
of the quadratic radius, which is given by $\langle r^2 \rangle \, =\, 6 \, a_1$ .

In summary, in this work we have used rational approximants as a tool for fitting the pion vector form factor. Because these approximants are capable of describing the region of large momentum, they may be better suited than polynomials for a description of the space-like data. As our results in Table 2 show, the errors achieved with these approximants are competitive with previous analyses existing in the literature.

\begin{table}[t]
\begin{center}
\begin{tabular}{|c|c|c|c|c|c|c|c|c|}
  \hline
   & $\langle r^2\rangle$   (fm$^2$)  &  $a_2$  (GeV$^{-4}$)     \\ \hline \hline
  This work &   $0.445\pm 0.002_{\mathrm{stat}}\pm 0.007_{\mathrm{syst}}$ &  $3.30\pm 0.03_{\mathrm{stat}}\pm 0.33_{\mathrm{syst}}$    \\
  CGL~\cite{Colangelo,ColangeloB}& $ 0.435\pm 0.005$  & ...   \\
  TY~\cite{Yndurain}  & $ 0.432\pm 0.001 $  & $ 3.84\pm 0.02$ \\
  BCT~\cite{op6-VFF} & $0.437\pm 0.016$  & $3.85\pm 0.60$ \\
  PP~\cite{Portoles} & $0.430\pm 0.012$  & $3.79\pm 0.04$ \\
   Lattice~\cite{lattice} & $0.418\pm 0.031$  & ... \\
  \hline
\end{tabular}
\caption{{\small Our results for the quadratic radius $\langle r^2\rangle$ and second derivative $a_2$ are
compared to other determinations~\cite{Colangelo,ColangeloB,Yndurain,op6-VFF,Portoles,lattice}. Our first error is
statistical. The second one is systematic, based on the analysis of the VFF model in section 2.}}
\label{table2}
\end{center}
\end{table}

\vspace{1cm}

\textbf{Acknowledgements}

We would like to thank G. Huber and H. Blok for their help
with the experimental data.
This work has been supported by
CICYT-FEDER-FPA2005-02211, SGR2005-00916, the Spanish Consolider-Ingenio 2010
Program CPAN (CSD2007-00042)
and by the EU Contract No. MRTN-CT-2006-035482, ``FLAVIAnet''.

\end{document}